\documentclass[10pt,twocolumn,a4paper]{article}

\usepackage[T1]{fontenc}
\usepackage[cp1250]{inputenc}
\usepackage{graphicx}
\usepackage{epstopdf}
\usepackage{lmodern}
\usepackage{booktabs}
\usepackage{rotating}

\usepackage{varioref}
\usepackage{amsmath,amssymb,amsfonts}

\usepackage[pdftex,breaklinks=true]{hyperref}
\hypersetup{
colorlinks=true,
linkcolor=black,          
citecolor=black,        
filecolor=black,      
urlcolor=black,           
bookmarks=true,
bookmarksnumbered=true,
pdfauthor={Piotr Bania},
pdfstartview=FitH,
pdfsubject={Securing The Kernel Via Static Binary Rewriting and Program Shepherding},
pdftitle={Securing The Kernel Via Static Binary Rewriting and Program Shepherding}
}

\usepackage{breakurl}
\usepackage{url}
\usepackage{algorithmic}
\usepackage{listings}
\usepackage{color}
\usepackage[ruled,vlined]{algorithm2e}

\title{Securing The Kernel via Static Binary Rewriting and Program Shepherding}
\author{Piotr Bania\\
\texttt{\href{http://www.piotrbania.com}{www.piotrbania.com}}}
\date{2011}

\begin{document}
\maketitle

\begin{abstract}

Recent Microsoft security bulletins show that kernel vulnerabilities are becoming more and more important security threats. Despite the pretty extensive security mitigations many of the kernel vulnerabilities are still exploitable. Successful kernel exploitation typically grants the attacker maximum privilege level and results in total machine compromise.

To protect against kernel exploitation, we have developed a tool which statically rewrites the Microsoft Windows kernel as well as other kernel level modules. Such rewritten binary files allow us to monitor control flow transfers during operating system execution. At this point we are able to detect whether selected control transfer flow is valid or should be considered as an attack attempt.

Our solution is especially directed towards preventing remote kernel exploitation attempts. Additionally, many of the local privilege escalation attacks are also blocked (also due to additional mitigation techniques we have implemented). Our tool was tested with Microsoft Windows XP, Windows Vista and Windows 7 (under both virtual and physical machines) on IA-32 compatible processors. Our apparatus is also completely standalone and does not require any third party software.
\end{abstract}

\section{Introduction}

Our tool uses the program shepherding technique~\cite{ProgramShep} (monitoring
control flow transfers). In our approach, however, we are not using a dynamic binary instrumentation engine, but instead all the selected binary files (Windows kernel and modules) are statically rewritten\footnote{Using dynamic binary instrumentation engines on kernel level code is surely harder to implement and more dangerous to use.}. By monitoring the control flow transfers we can ignore the complexities of various vulnerabilities and focus on preventing the execution of malicious code. Section~\ref{sec:design} describes the design of our tool.

\section{Design}
\label{sec:design}

Our apparatus consists of four main modules:

\begin{description}
    \item[Integration\footnote{In this paper we will refer to integration as a synonym of static binary code rewriting.}] Responsible for disassembling, analyzing and rewriting binary data{\footnote{In this paper integration process is limited to kernel and kernel mode components.}};
    \item[Monitoring] Responsible for filtering and monitoring the control flow transfers;
    \item[Configuration] Responsible for configuring, loading and unloading the monitoring module;
    \item[Installer] Responsible for replacing the rewritten kernel and necessary drivers.
\end{description}

The following subsections describe each module more thoroughly.

\subsection{Integration module}

Integration module is the most important and complex part of this project. It can be divided in two separate submodules: analyzer module and rewriting module.

\subsubsection{Analyzer module}

The analyzer module is responsible for disassembling and analyzing binary programs. In our case the binary programs are written in the Portable Executable (PE) file format.  The analyzer engine is a slightly modified version of the one used in our previous automated binary differential analysis project called AutoDiff~\cite{AutoDiff}. This component is completely standalone and does not rely on other disassembly engines, like the widely used commercial product IDA Pro~\cite{ida_pro_com}.

The analyzer provides structured information for the integration module. This includes the information about the instructions, basic blocks, functions and other important data. The analyzer engine also divides the analyzed code into two major categories:

\begin{description}
    \item[Solid code] Obtained in a result of recursive traversal disassembly and other heuristic techniques based on the Portable Executable file format characteristics;
    \item[Prospect code] Obtained in any other fashion, e.g., by using relocation information or by testing solid instruction operands.
\end{description}

Since our engine is using recursive traversal disassembly, dividing the analyzed code into those two categories helps us to avoid further code vs data misunderstandings. Additional heuristics mechanisms are also used, since the complexity of the executable binary files is often high. We also try to improve the code coverage by using the Microsoft symbol files. The main rule we have been using at this point is that it is better to confuse code as data than vice versa. This will be further explained in the next subsection.

As a result we achieve very good code coverage together with excellent overall performance (see Section~\ref{sec:int_performance}).

\subsubsection{Rewriting module}\label{int_module}

With the results gathered during the previous step, the rewriting module can perform the static code rewriting process. Basing on our previous experiences with static code rewriting like in project Aslan~\cite{aslan_website} or SpiderPig~\cite{spiderpig_project_website}, we have decided to use a more secure (stable) approach. In Aslan~\cite{aslan_website} the approach was to interfere with all the original instructions and data. This of course often required manual interaction because of the code vs data dilemma, which cannot be totally resolved by the static analysis. Such an approach is not really usable, as it cannot be fully automated. Thus, we have decided to modify and use SpiderPig~\cite{spiderpig_project_website} instead (it is easier and more secure to perform).

\paragraph{Code Integration Method}

Our tool can rewrite the binary files in two general ways. The first way is an non-invasive one, where the rewritten code is placed in a separate file. In other words, the original files are not modified. This approach requires the additional driver module to load the integrated code into the operating system and patch all the necessary original modules in virtual memory. This approach is considered to be more secure (in terms of stability) since the original files are not changed. The second way (which is currently implemented and used in our tool) is more invasive since it modifies (rebuilds) the original Portable Executable files. In other words, the operating system boots up with an already modified kernel and selected kernel modules. We will describe our code integration algorithm in reference to the currently used method (invasive).

One of our initial assumptions in the process of static binary rewriting was to preserve the original file structure in such a way that the original code and data offsets are not changed. This step is essential for increasing the stability of the rewritten code and avoiding other problems like the already mentioned code vs data dilemma. The newly generated code which includes all the original functions is attached to the end of the original file. Typically the new code resides in the relocation section of the original Portable Executable file. The destination section is expanded and modified in order to handle executable and non-pageable code. Code is rewritten such that all the original functions are instrumented (this will be described in the next paragraph) and the overall functionality is preserved. Rewritten functions also do not contain shared basic blocks\footnote{This is done especially for some protection methods we would like to implement in the future~\cite{PaxFuture, RopArticlePB}.}. The binary rewriting process can be divided into three phases:

\begin{description}
    \item[Instrumentation] Responsible for adding instrumentation code and expanding original control transfer instructions;
    \item[Calculation] Responsible for allocating new relative virtual addresses for the rewritten basic blocks and for generating new relocation entries;
    \item[Repairing] Responsible for repairing relative offsets of control transfer instructions.
\end{description}

In the {\emph{Instrumentation phase}} we have two primary objectives. First one is to expand all the short conditional and unconditional jumps in order to avoid further problems with fixing the relative offsets in the repairing phase. Second one is strict requirement of the program shepherding method. In this case we add instrumentation for every instruction that is either a {\tt{CALL}} or indirect {\tt{JMP}} or a {\tt{RET}} instruction. Instrumentation is performed in such a way that the filtering procedure is executed before the original instructions. This gives us the possibility to deny the control transfers into the malicious memory region. It is worth noticing that we are not instrumenting {\tt{CALL}} or {\tt{JMP}} indirect instructions which refer to imported API functions. Firstly because the import address table is typically read-only (presents low security threat), and secondly due to performance reasons.

The {\emph{Calculation phase}} is responsible for the calculation of new relative virtual addresses of the newly generated basic blocks. Additionally it is also needed for creating new relocation entries for the generated code. If an original instruction consists of an operand (either immediate or memory immediate) that has a relocation entry, the rewritten instruction requires the corresponding relocation entry as well.

The {\emph{Repairing phase}} is necessary for fixing the offsets of the control transfer instructions that use relative operands. This is essential for keeping the rewritten correct stable and to ensure that the execution flow will remain in the rewritten code. At this point we don't need to repair other control transfer instructions since we assume the control will be given back due to the emitted function hooks (see next paragraph).

When all the phases are finished the engine is almost ready to produce a new Portable Executable file that contains the rewritten code. However, there is one essential step yet to perform. In order to transfer the execution from original code to the rewritten code, the engine must patch original functions and redirect them to the corresponding generated ones. This is achieved by emitting the {\tt{JMP}} relative instruction at the original function prologue. This is not so easy to perform since the first basic block of the original function may be smaller than the 5 bytes required for the patch. Additionally, we may confuse code with data and patch some essential kernel structure which in the end will lead to a system crash. 

In order to do this safely, our engine decides whether the original function can be patched or not. The decision is based on a few tests. First of all we check the size of first basic block. If it is not large enough original function remains unpatched. For the functions marked as {\textbf{prospect code}} we apply some additional tests. For example, we don't patch functions that consist only of one basic block. We also test if the selected function is entirely built from {\tt{ASCII}} or {\tt{Unicode}} characters. As it was mentioned earlier we prefer to redirect less functions and cause no stability issues than vice versa. It is also worth mentioning that the Windows kernel uses self-modifying code at certain places, forcing us to use additional checking mechanisms to address such problems. Even though we don't redirect all the original functions, our tests showed that it is still more than enough for our solution to work successfully. After the original functions are patched, the engine checks the relocation entries one more time in order to make sure none of them overlaps with the 5 byte relative {\tt{JMP}} instruction. If such relocation entry is found it is simply erased. Lastly the Portable Executable headers are fixed. When all the issues are resolved the new PE file is emitted.

It is important to note that the static binary rewriting process may be performed on a different (remote) machine.

\subsection{Monitoring module}

Monitoring module is developed as a device driver. It acts as a server for the configuration module. It is also responsible for filtering and monitoring the control transfers caused by instrumented instructions. The filtering method is described in Section~\ref{exploit_detection}. Monitoring module contains a memory map structure that includes the information about the currently loaded kernel modules. This memory map is updated after selected device driver is loaded to or unloaded from the kernel memory. It is worth noting that instrumented instructions in the rewritten binary files do not execute the filtering procedure before the monitoring module is not fully initialized. Monitoring module is also responsible for blocking most of the local privilege escalation exploits by utilizing a very simple but yet effective technique (described in the Section~\ref{local_mitigation}).

\subsubsection{Mitigation technique for local privilege escalation attacks}\label{local_mitigation}

Most of the local privilege escalation exploits use the {\tt{NtQuerySystemInformation}}\footnote{{\tt{EnumDeviceDrivers}} is also used however this API function is just a wrapper for {\tt{NtQuerySystemInformation}}.}  function to gather the base addresses where the kernel modules are mapped to. On Microsoft Windows systems, device drivers are mapped to different memory addresses each run. Thus, getting their base addresses is very often essential for such exploits to work correctly. Our solution hooks the {\tt{NtQuerySystemInformation}} function and denies all user-mode requests where the {\tt{SystemModuleInformation}} class is passed as parameter. According to our tests this method does not influence the operating system stability\footnote{Our tests were performed on a default instalations of Microsoft Windows systems.}. At this point the only way for the attacker to succeed is to find the base address using some other method (for example by using another vulnerability).

\subsubsection{Detecting exploitation attempts}\label{exploit_detection}

In order to detect attack attempts, the filtering (monitoring) procedure needs to decide whether selected control transfer is valid or not. Filtering procedure must be fast enough to not cause any major slowdown of the operating system. Our tool uses the memory map structure (mentioned in the previous section) which contains information about the currently loaded device drivers. This information is divided into fast memory page lookup entries which provide the characteristics of the selected page. For example it shows whether the page is writable, executable or whether it is a part of the kernel or any other loaded module. At this point our security policy is very simple. We mark all the control transfers to pages that are not executable and are not the part of kernel or other modules as attack attempts. Since we don't store the information about the userland modules, all control transfers from kernel to usermode space are automatically marked as forbidden.

\subsubsection{Reaction to attack}

When an attack is detected there are two available reaction options in the monitoring module. First one is to log the attack in a specified file and continue the execution; this is especially useful for honeypot-like systems. The second choice provides the attack logging feature together with immediate system shutdown. It is important to notice that we are operating in the kernel mode. Therefore there is no single task we can securely terminate in comparison to user mode solutions where such action can be typically safely performed.

\subsection{Configuration module}

The Configuration module loads the monitoring module driver and creates initial memory map for all of the currently loaded kernel modules. It also provides additional information required by the monitoring module. Our policy allows only one configuration attempt after system start. This is done in order to block other potentially malicious configuration requests from the attacker.

\subsection{Installer module}

This module currently consists of batch scripts and programs that allow one to modify the Microsoft kernel and selected device drivers. On Windows XP we are using a {\tt{WINLOGON.EXE}} thread injection method to disable the Windows File Protection. This is achieved by executing the  {\tt{SfcTerminateWatcherThread}} API from {\tt{SFC\_OS.DLL}} library. On Microsoft Windows Vista and Windows 7 we need to take file ownership and grant full access control permissions to ourselves. Additionally {\tt{WINLOAD.EXE}} is copied and patched in order to allow the execution of modified Windows kernel. We are currently working on fully automating described tasks (together with easy file recovery) for all Windows operating systems.

\section{Experimental results}

This section is divided into two subsections. First one describes the results we have obtained while testing versus selected Windows exploits. Second one shows the performance impact.

\subsection{Effectiveness}

We have tested our solution with a few publicly available exploits. Due to small number of publicly released exploits (especially remote ones) targeting kernel and kernel modules, our tests are currently limited. Obtained results are presented below:

\begin{itemize}
    \item {\textbf{CVE-2009-3103}} (Microsoft Windows SMB2 '{\tt{Smb2ValidateProviderCallback}}' Remote Code Execution Vulnerability) \cite{SMB2CVE, SMB2Exploit} \newline
{\newline\noindent}This vulnerability allows remote attackers to execute arbitrary code with system privileges. Reliable and publicly known exploit for this issue (see \cite{SMB2Exploit} for technical details on the exploitation process) firstly generates a so called trampoline which is located at fixed BIOS/HAL memory region (which is by default readable, writeable, and executable). After the trampoline is ready the execution is thrown to it using a {\tt{CALL EAX}} instruction (located in {\tt{srv2.sys}} module). Our solution is able to detect and block this attack before the generated trampoline is executed.

    \item {\textbf{CVE-2010-2743}} (Microsoft Windows {\tt{win32k.sys}} Keyboard Layout Vulnerability) \cite{StuxnetCVE, StuxnetVUPEN, StuxnetRUBEN} \newline
{\newline\noindent}This is one of the local vulnerabilities exploited by the Stuxnet worm \cite{StuxnetWIKI} in order to elevate privileges. Keyboard layout vulnerability is caused by {\tt{win32k!xxxKENLSProcs}} function that do not properly perform indexing of a function-pointer table during the loading of keyboard layouts from disk. Malicious code in this case is executed by the {\tt{CALL \_aNLSVKFProc[ecx*4]}} instruction. Due to forged index value the code flow is redirected to {\tt{0x60636261}}. Contents of the memory located at this specific address can be controlled by the attacker. Tests showed that our tool detects and prevents successful exploitation of this vulnerability.
\end{itemize}

Most of the others local privilege escalation exploits are unable to work at the very first stage. This is caused by the mitigation that was described in the section \ref{local_mitigation}. \newline
{\newline\noindent}As a ``side effect'' our apparatus is able to detect some cases of hidden malicious code that operates at the kernel level (like rootkits or bootkits).

\subsection{Performance}

Following section presents the performance results of our tool. It is divided into two subsections. Where first one describes the performance of the integration module and the second one focuses on testing the protected operated system itself.

\subsubsection{Integration performance}\label{sec:int_performance}

Table~\ref{tab:tabela_int} presents the results obtained by integrating original Microsoft Windows 7 files (listed below). This test was performed on T3400 2.16Ghz (Core2) notebook machine with 2.46GB of RAM. Modules (files) presented in the table were chosen specifically due to potential security threats they create. This does not mean our tool is not able to protect other modules.\newline
{\newline}The legend for Table~\ref{tab:tabela_int} is as follows:
\begin{itemize}
    \item Size$_{org}$ - original file size,
    \item Size$_{int}$ - file size after integration,
    \item T$_{disasm}$ - time required for disassembling the selected file\footnote{Does not include time required for downloading symbol file etc.},
    \item T$_{basicblock}$ - time required for creating basic blocks from the disassembly information
    \item T$_{int}$ - time required for instrumenting, repairing and generating new code\footnote{Does not include time required for emitting the PE file.}.
\end{itemize}

Presented results indicate that the integration process is more than satisfactory in terms of speed and memory usage. Results also show that typically newly generated files are twice as large as the original ones. This is natural considering the modifications we have applied. As it was mentioned earlier (see Section~\ref{int_module}), our integration engine may also work on external machines. This gives user the opportunity to perform the binary rewriting process remotely.

\begin{table*}[!ht]
  \centering
  \caption{Static binary rewriting performance depending on a various files.}
    \begin{tabular}{|p{1.75cm}|p{1.3cm}|p{1.3cm}|p{1.4cm}|p{1.45cm}|p{1.60cm}|p{1.55cm}|p{1.7cm}|p{1.35cm}|}
    \addlinespace
    \toprule
    File & Size$_{org}$ [MB]& Size$_{int}$ [MB] & T$_{disasm}$ [sec] & T$_{basicblock}$ [sec] & Instructions [\#] & Basic blocks [\#] & Memory usage [MB] & T$_{int}$ [sec]\\
    \midrule
    afd.sys & 0.323242 & 0.628418 & 0.089667 & 0.064936 & 80506 & 20250 & 19.222656 & 0.077218 \\
    \hline
    http.sys & 0.489258 & 0.942383 & 0.135878 & 0.132798 & 120746 & 30186 & 27.746094 & 0.135387 \\
    \hline
    mrxsmb.sys & 0.117676 & 0.241699 & 0.042676 & 0.042676 & 31537 & 7868  & 9.316406 & 0.033276 \\
    \hline
    ndis.sys & 0.677795  & 1.339844 & 0.198545 & 0.320212 & 168001 & 42421 & 37.218750 & 0.169197 \\
    \hline
    ndistapi.sys & 0.020020 & 0.035645  & 0.004350 & 0.002852 & 4289  & 1062  & 5.316406 & 0.003737 \\
    \hline
    ndproxy.sys & 0.045898 & 0.089844 & 0.016023 & 0.008206 & 11744 & 2889  & 5.593750  & 0.011264  \\
    \hline
    netbios.sys & 0.034668  & 0.068848 & 0.009053 & 0.007547 & 9142  & 2538  & 5.304688 & 0.008585 \\
    \hline
    netbt.sys & 0.179199  & 0.377441 & 0.057341 & 0.048043 & 53761 & 13383 & 13.437500 & 0.054077 \\
    \hline
    ntkrnlpa.exe & 3.773804 & 8.202148 & 1.332446 & 1.082007 & 998898 & 259320 & 221.421875 & 5.808912 \\
    \hline
    ntoskrnl.exe & 3.721069 & 8.085449 & 1.310291 & 1.075641 & 982377 & 256603 & 218.273438 & 6.477548 \\
    \hline
    smb.sys & 0.067871 & 0.133301 & 0.018859 & 0.013599 & 17767 & 4387  & 12.523438  & 0.017782 \\
    \hline
    srv2.sys & 0.295410 & 0.520508 & 0.060904 & 0.051449 & 58071 & 14643 & 16.945313 & 0.057214 \\
    \hline
    srv.sys & 0.296875 & 0.639160 & 0.089117  & 0.075273 & 82425 & 21662 & 21.957031  & 0.092641 \\
    \hline
    tcpip.sys & 1.226440  & 2.475098 & 0.411125 & 0.499790 & 335331 & 79187 & 72.378906 & 0.344304 \\
    \hline
    tdi.sys & 0.020020 & 0.041016 & 0.004489 & 0.002879 & 4348  & 1110  & 13.511719 & 0.004962 \\
    \hline
    win32k.sys & 2.223145 & 4.992676 & 0.952298 & 0.981985 & 673535 & 175847 & 147.082031 & 4.389956 \\
    \bottomrule
    \end{tabular}%
 \label{tab:tabela_int}%
\end{table*}%

\subsubsection{System performance}


In the system performance testing we have used our own custom benchmarking tool and also two other solutions for Microsoft Windows systems (NovaBench version 3~\cite{BenchmarkNova} and PerformanceTest 7~ \cite{BenchmarkPassmark} evaluation version).

\paragraph{Benchmarked machine configuration:}Intel Core2 Q9550 2.83GHz; 3327 MB RAM; ATI Radeon HD 5870; Windows 7 (32-bit).\newline

Our benchmark results are presented in Table \ref{tab:nasz_benchmark}, NovaBench results are presented in Table \ref{tab:novabench}. PerformanceTest benchmark results are presented in Table~\ref{tab:passmark}.\newline

The legend for Table~\ref{tab:nasz_benchmark} is as follows:
\begin{itemize}
    \item P$_{1}$ - protected machine (full instrumentation),
    \item Ps$_{1}$ - slowdown (P$_{1}$ versus native configuration),
    \item P$_{2}$ - protected machine (skipped {\tt{RET}} instruction instrumentation in {\tt{win32k.sys}} - otherwise full instrumentation),
    \item Ps$_{2}$ - slowdown (P$_{2}$ versus native configuration).
\end{itemize}

\begin{table}[!h]
  \centering
  \caption{Custom benchmark results on native and protected systems.}
    \begin{tabular}{|p{1.6cm}|p{0.9cm}|p{0.6cm}|p{0.7cm}|p{0.6cm}|p{1cm}|}
    \addlinespace
    \toprule
    Test  & Native [s] & P$_{1}$ [s] & Ps$_{1}$ [\%] & P$_{2}$ [s] & Ps$_{2}$ [\%] \\
    \midrule
    Process & 3.44  & 5.53  & 60.68 & 4.22  & 22.43 \\
    \hline
    Write File & 7.32  & 7.73  & 5.62  & 7.57  & 3.39 \\
    \hline
    Read File & 2.27  & 2.28  & 0.49  & 1.98  & -12.88 \\
    \hline
    Memory & 0.65  & 0.73  & 12.29 & 0.71  & 9.84 \\
    \bottomrule
    \end{tabular}%
  \label{tab:nasz_benchmark}%
\end{table}%

\begin{table}[htbp]
  \centering
  \caption{NovaBench 3 results on native and protected system (higher number the better).}
    \begin{tabular}{|p{3.5cm}|p{1.5cm}|p{1.6cm}|}
    \addlinespace
    \toprule
    Test  &  Native system & Protected system \\
    \midrule
    RAM Speed (MB/s) & 4203  & 4178 \\
    \hline
    Floating Point (ops/s) & 102153392 & 102077740 \\
    \hline
    Integer (ops/s) & 333407424 & 333462560 \\
    \hline
    MD5 Hashes (gen/s) & 929197 & 924222 \\
    \hline
    CPU Score & 403   & 403 \\
    \hline
    Graphics Tests Score & 494   & 481 \\
    \hline
    Drive Write Speed (MB/s) & 79    & 70 \\
    \hline
    Hardware Tests Score & 30    & 28 \\
    \hline
    NovaBench Score & 1030  & 1014 \\
    \bottomrule
    \end{tabular}%
  \label{tab:novabench}%
\end{table}%

\begin{table}[htbp]
  \centering
  \caption{PerformanceTest 7 results on native and protected system (higher number the better).}
    \begin{tabular}{|p{4.4cm}|p{1cm}|p{1.28cm}|}
    \addlinespace
    \toprule
    Test  &  Native system & Protected system \\
    \midrule
    CPU - Integer Math & 547.0 & 543.4 \\
   \hline
    CPU - Floating Point Math & 2129.9 & 2131.4 \\
   \hline
    CPU - Find Prime Numbers & 1157.6 & 1155.6 \\
   \hline
    CPU - Multimedia Instructions & 9.6   & 9.6 \\
\hline
    CPU - Compression & 6306.5 & 6310.5 \\
   \hline
    CPU - Encryption & 17.9  & 17.9 \\
   \hline
    CPU - Physics & 314.3 & 313.1 \\
   \hline
    CPU - String Sorting & 3634.2 & 3642.8 \\
   \hline
    Graph2D - Solid Vectors & 3.9   & 2.5 \\
   \hline
    Graph2D - Transparent Vectors & 3.8   & 2.4 \\
   \hline
    Graph2D - Complex Vectors & 93.0  & 50.5 \\
   \hline
    Graph2D - Fonts and Text & 113.1 & 63.4 \\
   \hline
    Graph2D - Windows Interface & 68.7  & 36.3 \\
   \hline
    Graph2D - Image Filters & 415.9 & 408.6 \\
   \hline
    Graph2D - Image Rendering & 289.3 & 272.2 \\
   \hline
    Graph3D - Simple & 3568.4 & 2963.7 \\
   \hline
    Graph3D - Medium & 843.9 & 843.3 \\
   \hline
    Graph3D - Complex & 81.8  & 77.7 \\
   \hline
    Graph3D - DirectX 10 & 55.8  & 55.7 \\
   \hline
    Memory - Small Block Alloc & 2767.4 & 2747.3 \\
   \hline
    Memory - Read Cached & 2195.2 & 2195.0 \\
   \hline
    Memory - Read Uncached & 2027.7 & 2021.7 \\
   \hline
    Memory - Write & 2095.3 & 2082.5 \\
   \hline
    Memory - Large RAM & 1177.9 & 1076.1 \\
   \hline
    Disk - Sequential Read & 86.2  & 85.6 \\
   \hline
    Disk - Sequential Write & 85.7  & 90.2 \\
   \hline
    Disk - Random Seek + RW & 3.5   & 3.4 \\
   \hline
    CPU Mark & 3727.8 & 3726.1 \\
   \hline
    2D Graphics Mark & 475.2 & 318.5 \\
   \hline
    Memory Mark & 872.1 & 843.9 \\
   \hline
    Disk Mark & 634.2 & 648.3 \\
   \hline
    3D Graphics Mark & 2749.8 & 2584.6 \\
   \hline
    PassMark Rating & 1612.4 & 1366.3 \\
    \bottomrule
    \end{tabular}%
  \label{tab:passmark}%
\end{table}%

Each benchmark program tend to produce results that vary during each system run. Our custom benchmarking tool executes four types of tests. The {\textit{Process Test}} works by creating 100 instances of {\tt{calc.exe}} program. The time is measured until all of the created processes are fully initialized. The {\textit{Write File Test}} creates 100 10MB files and fills them with constant data. The  {\textit{Read File Test}} works in analogical way. In both cases time is measured until all files are processed. Last test ({\textit{Memory Test}}) allocates (commits) 100 memory regions each 100MB wide, fills them with constant data and finally frees the committed regions. Time is measured until this process is finished for all the regions. Each test was performed 5 times for both protected and unprotected configurations (5 samples were taken for each configuration). The arithmetic mean of the results was used in the comparison process.

Our benchmark showed that the largest performance impact was observed in the {\textit{Process Test}}. The slowdown in this case was approximately 60\% (P$_1$ case). As it was mentioned already {\textit{Process Test}} rest on creating 100 {\tt{calc.exe}} processes. In case of the P$_2$ configuration the slowdown was limited to about 22\%. The only difference between P$_1$ and P$_2$ configuration was that P$_2$ configuration skipped the {\tt{RET}} instrumentation in the {\tt{win32k.sys}} module. This device driver is a major component of the Windows GUI subsystem. Since {\tt{calc.exe}} is a GUI process we assume that the negative performance impact was specifically caused by the graphical interface initialization for this process. Likewise in the Passmark benchmark, protected system causes some negative performance effect on 2D graphics tests (Table \ref{tab:passmark}). This is also especially caused by instrumented {\tt{win32k.sys}} module. We have made another test. We have skipped the {\tt{RET}} instruction instrumentation again and the performance results were significantly improved. According to the rest of the benchmarks results there is also a slight overall performance impact regarding memory allocation, disk read, disk write operations (albeit it is not as prominent as the impact on 2D graphics performance). Furthermore our benchmark (Table \ref{tab:nasz_benchmark}) indicated that in the {\textit{Read File Test}} (P$_2$ configuration) our performance was almost 13\% better in comparison to the native machine. However this may be caused by the Windows file caching mechanisms. 
{\newline\indent}Future versions of the engine should address the performance issues. We have already presented optimization ideas (section \ref{sec:future_work}) that should significantly improve the overall performance in the future. We also plan improve the effectiveness of our solution without skipping the {\tt{RET}} instrumentation. Additionally it is also worth noticing that performance slowdowns presented in our benchmarks are often not widely manifested.

\section{Future Work}\label{sec:future_work}

Currently our system is in proof-of-concept state. We are planning to extend it into two general directions. In order to improve performance we want to instrument only those areas that represent a high security threat (right now we are instrumenting all of them without evaluating the security threat). In the case of {\tt{RET}} instructions this work is easier since Windows supports kernel stack cookies (starting from Microsoft Windows XP). Therefore such instructions can be left not instrumented. Additional work should be done in order to limit the number of control transfers to the callback functions --- this should improve the efficiency of the CPU instruction cache. Due to nature of Microsoft Windows updates that usually ship every month we need to monitor whether selected module was changed by the update or not\footnote{Typical Microsoft updates involve relatively small number of kernel modules updates.}. If so it needs to be rewritten again. Easy file recovery method is also in plans. 
{\newline\indent}Another direction involves expanding the security policy. In this case the one presented in~\cite{ProgramShep} (or rather some parts of it we have found most interesting) can be accommodated. For example allowing indirect calls to target only valid entry points within other modules. Another idea is to statically predict whether selected function is used only locally (within the selected module) and design the security policy for this function accordingly. For example (in case of function return) all control transfers to places outside the current module from this function epilogue will be forbidden. Similar ideas can apply to other instructions that cause indirect control flow transfers. Additionally to prevent return-oriented programming (ROP) attacks~\cite{RopPaper, RopPresent, RopDINO} we can emit magic bytes (a key) for every function call and check it when the function returns~\cite{PaxFuture, RopArticlePB}. Further mitigations for return-oriented programming attacks may involve generating different file variants per machine and eradicating original code where possible. Regarding local attacks it may be beneficial to disallow an unprivileged user from accessing system modules (kernel and drivers). This includes any type of access. 
{\newline\indent}In order to block potential clobbering of the memory map by the attacker it may be beneficial to store it at read-only memory region. This action would require to change the memory rights to writable only when kernel module is loaded or unloaded from the memory. Thus it shouldn't cause any serious performance impact. 
{\newline\indent}It is possible to port our solution to {\emph{x86-64 architectures}}. However it is important to notice that on x64 platforms Microsoft has introduced a new feature, called PatchGuard that is intended to prevent both malicious software and third-party vendors from modifying certain critical operating system structures. Even though this security mechanism is bypassable~\cite{PatchGuardBypass, PatchGuardBypass2} one may wonder whether is it worth to disable one security feature for the sake of another.

\section*{Acknowledgments}

Author would like to thank Brad Spengler and Matt Miller for helping with this article.

\section{Conclusion}

This paper presents a method for securing the kernel of the operating system (in this case Microsoft Windows). Our engine uses static binary rewriting and code instrumentation techniques in order to monitor the control flow. We have shown that our protection is capable of detecting and blocking both remote (especially) and local attacks. Our solution, however, does not prevent against exploits that overwrite sensitive data and it also does not protect against vulnerabilities in 3rd party kernel modules (besides the technique presented in section \ref{local_mitigation}). 

We have also described techniques and ideas that may be implemented in the future. Performance impact together with operating system benchmarks was also presented. We believe that currently our solution provides a unique security solution for the Microsoft Windows kernel and does not require any special hardware features. Of course this mechanism cannot solve every security problem completely but it does make kernel exploitation much harder.

\newpage
\bibliographystyle{plain}
\bibliography{F:/latex_sheets/spiderpig_thesis/spiderpig/bibliografia}
\end{document}